\documentclass[twocolumn,pra,showpacs,aps,superscriptaddress, reprint]{revtex4-1}

\usepackage[english]{babel}
\usepackage[ansinew]{inputenc}

\usepackage[dvipsnames]{xcolor}

\usepackage{times}
\usepackage{graphicx}
\usepackage{graphics}
\usepackage{amsmath}
\usepackage{amsfonts}
\usepackage{amssymb}
\usepackage{dsfont}
\usepackage{epstopdf}
\usepackage{makeidx}
\usepackage{subfigure}
\usepackage{color}
\usepackage{pgf}
\usepackage{bm}
\usepackage{tikz} 
\usepackage[normalem]{ulem}
\usepackage{units}

\newcommand{\be}{\begin{equation}}
\newcommand{\ee}{\end{equation}}
\newcommand{\bea}{\begin{eqnarray}}
\newcommand{\eea}{\end{eqnarray}}

\makeindex
\begin{document}
\title{Purcell effect at metal-insulator transitions}
\author{D. Szilard}
\affiliation{Instituto de F\'{\i}sica, Universidade Federal do Rio de Janeiro,
Caixa Postal 68528, Rio de Janeiro 21941-972, RJ, Brazil}
\author{W. J. M. Kort-Kamp}
\affiliation{Theoretical Division and Center for Nonlinear Studies, MS B258, Los Alamos
National Laboratory, Los Alamos, New Mexico 87545, United States}
\author{F. S. S. Rosa}
\affiliation{Instituto de F\'{\i}sica, Universidade Federal do Rio de Janeiro,
Caixa Postal 68528, Rio de Janeiro 21941-972, RJ, Brazil}
\author{F. A. Pinheiro}
\affiliation{Instituto de F\'{\i}sica, Universidade Federal do Rio de Janeiro,
Caixa Postal 68528, Rio de Janeiro 21941-972, RJ, Brazil}
\author{C. Farina}
\affiliation{Instituto de F\'{\i}sica, Universidade Federal do Rio de Janeiro,
Caixa Postal 68528, Rio de Janeiro 21941-972, RJ, Brazil}
\date{\today}

\begin{abstract}
We investigate the spontaneous emission rate of a two-level quantum emitter next to a composite medium made of randomly distributed  metallic inclusions embedded in a dielectric host matrix. In the near-field, the Purcell factor can be enhanced by two-orders of magnitude relative to the case of an homogenous metallic medium, and reaches its maximum precisely at the insulator-metal transition. By unveiling the role of the decay pathways on the emitter's lifetime, we demonstrate that, close to the percolation threshold, the radiation emission process is dictated by electromagnetic absorption in the heterogeneous medium. We show that our findings are robust against change in material properties, shape of inclusions,  and apply for different effective medium theories as well as for a wide range of transition frequencies. 
\end{abstract}
\maketitle

\vspace{0.1in}

\section{Introduction}
In cavity quantum electrodynamics, spontaneous emission (SE) is a pivotal
example of energy transfer from an excited quantum emitter (atom, molecule, or 
quantum dot) into its environment. As first predicted by 
Purcell~\cite{Purcell-46} and later experimentally confirmed by Drexhage {\it et 
al.}~\cite{DrexhageEtAl-68}, the environment exerts a crucial influence on the 
emitters' decay. Indeed, the presence of objects in the system allow for new 
energy relaxation channels ({\it e. g.} plasmonic excitations) that can strongly 
affect the emitter's lifetime~\cite{novotnybook}. The influence of the environment on the emitter's radiative properties characterizes the Purcell effect and is 
quantified by the local density of optical states (LDOS). Typically, 
a modification of the LDOS involves either 
changing the environment geometry or its material properties 
~\cite{joulain2003}.

The last decade has witnessed an increasing research effort towards the control 
of SE rate due to the notable progresses in near-field optics, plasmonics, and 
metamaterials. Advances in nano-optics have not only allowed the 
improvement of the spectroscopical resolution of 
molecules in complex environments~\cite{betzig1993}, but have also led to the 
use of nanometric objects (e.g. nanoparticles and nanotips) that modify the 
lifetime and enhance the fluorescence of single 
molecules~\cite{bian1995,sanchez1999}. The advent of plasmonic devices and 
metamaterials has also opened new possibilities for tailoring the SE rate. 
Indeed, the local field enhancement due to excitation of plasmonic resonances 
has been explored in several applications, such as the surface-enhanced Raman 
scattering~\cite{jackson2004,wei2008,li2010}, and the modification of two-level 
atom resonance fluorescence~\cite{klimov2012}. In addition, photonic 
crystals~\cite{lodahl2004}, optical cavities~\cite{raimond2001},  metallic 
nanostructures~\cite{novotny2011, smith2014}, plasmonic 
cloaks~\cite{kortkamp2013,kortkampJOSA}, hyperbolic 
metamaterials~\cite{cortes2012} and 
negative index materials~\cite{klimov2002} are some examples of systems 
in which the LDOS and SE rate are dramatically affected by unusual 
photonic properties of the environment. Besides, gated and magnetic field biased 
graphene correspond to systems where active control of the Purcell effect can be implemented~\cite{tielrooij2015, 
kortkamp2015}.  However, in most of previous examples the modification of the 
LDOS involves sophisticated nano-fabrication techniques and/or complex 
nanostructures.  

In the present paper, we propose an alternative material platform, of easy 
fabrication, to tailor and control the SE of quantum emitters, namely, composite 
media. Our study is motivated by recent experimental observations that the SE is modified in the presence of 
metallic, semicontinuous 
media~\cite{krachmalnicoff2010,sapienza2011,nakamura2012}. Specifically, we have 
investigated the SE rate of a two-level atom in the vicinities of a 
semi-infinite medium composed of metallic inclusions, with various shapes and 
concentrations, embedded in a dielectric host medium. Applying different 
homogenization techniques (Bruggeman~\cite{Choy1999} and 
Lagarkov-Sarychev~\cite{Lagarkov1996}), we demonstrate that the SE rate is 
remarkably enhanced in composite media  in relation to the case where 
homogeneous media are considered. In particular, we show that SE rate is maximal 
precisely at the insulator-metal transition (percolation threshold). We 
demonstrate that these results are independent of the shape and material of the 
inclusions, and are valid for a broad range emission wavelengths. Altogether 
our findings suggest that composite media could be exploited in the design of novel, 
versatile materials in applications involving the radiative properties of light emitters. 

The paper is organized as follows. In Sec. II we present the employed methodology and the effective medium theory used to model the effective electric permittivity of the composite medium. In Sec. III we present our main results and the related discussions, while in Sec. IV we summarize the results and conclude.     

\section{Methodology}
\subsection{ The Purcell Effect }

\begin{figure}
\centering
\includegraphics[scale=0.42]{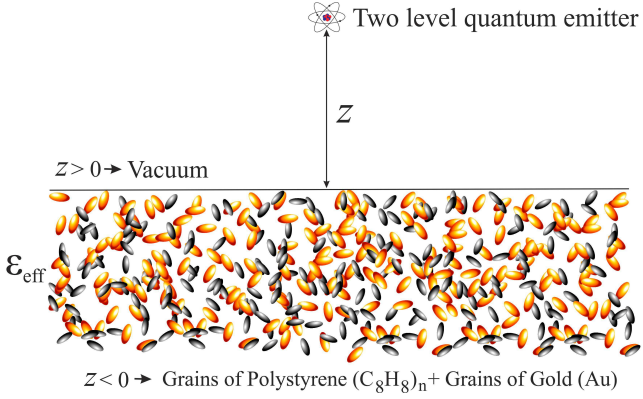}
\caption{Schematic view of the system under investigation: a two-level emitter at a distance $z$ of  a half-space composed of metallic (gold) inclusions embedded in a dielectric (polystyrene) host matrix.}
\label{Figure1}
\end{figure}

Let us consider a two-level emitter at a distance $z$ of a semi-infinite medium composed of randomly distributed metallic inclusions (electric permittivity $\varepsilon_{i}$) embedded in a dielectric host matrix (electric permittivity $\varepsilon_{hm}$), as shown in Fig.~\ref{Figure1}. In the presence of an arbitrary environment, the SE rate of a two-level atom reads \cite{novotnybook}

\begin{equation}
		\Gamma_{21} = \dfrac{6\pi c}{\omega_0} \, \Gamma^{(0)} \, \textrm{Im} \{ {\bf n} \cdot {\mathbb G} ({\bf r}, {\bf r}; \omega_0 ) \cdot {\bf n} \}\, ,
\label{TaxaEmissaoEspontanea2}
\end{equation}
where  $\Gamma^{(0)} = \omega_0^3 |{\bf d}_{21}|^2/3 \pi \hbar \varepsilon_0 
c^3$ is the free-space SE rate,  $\omega_0 = k_0 c$ is the transition frequency, 
${\bf d}_{21} $ the  emitter's transition electric dipole moment, ${\bf n} = 
{\bf d}_{21}/|{\bf d}_{21} |$, and $\mathbb{G} ({\bf r}, {\bf r}'; \omega)$ is 
the dyadic Green function of the system. The influence of the surrounding bodies on the 
emitter's lifetime is coded into  $\mathbb{G} ({\bf r}, {\bf r}'; \omega)$, 
which satisfies 

\begin{equation}
	\nabla \times \nabla \times \mathbb{G} ({\bf r}, {\bf r^{\prime}}; \omega ) - \dfrac{\omega^2}{c^2} \, \mathbb{G} ({\bf r}, {\bf r^{\prime}}; \omega) = \mathbb{I} \delta ({\bf r}, {\bf r^{\prime}}).
\label{FuncaoGreen}
\end{equation}

In free-space the dyadic Green function can be cast as\linebreak (for $z> z'$)
\begin{equation}
\label{FreeSpaceGreenFunction}
\mathbb{G}^{\textrm{(0)}}({\bf r}, {\bf r'}; \omega) = \dfrac{i}{2} \int  \dfrac{\mathbb{M} e^{i[{\bf k}_{\parallel}\cdot ({\bf r} - {\bf r}') + {k_z}_0(z-z')]}}{{k_{z}}_0}\dfrac{d^2{\bf k}_{\parallel}}{(2\pi)^2}\, ,
\end{equation}
where $k_{z0} = \sqrt{k_0^2-k_{\parallel}^2}$ and $\mathbb{M}$ is given by
\begin{equation}
\mathbb{M} = \mbox{{\mathversion{bold}${\epsilon}$}}_{\textrm{TE}}^{+} \otimes  \mbox{{\mathversion{bold}${\epsilon}$}}_{\textrm{TE}}^{+} +  \mbox{{\mathversion{bold}${\epsilon}$}}_{\textrm{TM}}^{+} \otimes  \mbox{{\mathversion{bold}${\epsilon}$}}_{\textrm{TM}}^{+}\, ,
\end{equation}
with the TE- and TM-polarization vectors defined as
\begin{eqnarray}
\mbox{{\mathversion{bold}${\epsilon}$}}_{\textrm{TE}}^{\pm} = \dfrac{-k_y{\bf \hat{x}} + k_x {\bf \hat{y}}}{k_{\parallel}}\, , 
\ \  \textrm{and} \ \  \mbox{{\mathversion{bold}${\epsilon}$}}_{\textrm{TM}}^{\pm} = \dfrac{\pm{k_z}_0{\bf k}_{\parallel}- k_{\parallel}^2{\bf \hat{z}}}{k_{\parallel}k_0}.
\end{eqnarray}
Note that these vectors are orthogonal, but they are normalized only for propagating modes ($k_{\parallel} < k_0$). Substituting Eq. (\ref{FreeSpaceGreenFunction}) into (\ref{TaxaEmissaoEspontanea2}) and using that at the coincidence  the only non vanishing components of $\mathbb{G}^{(0)}$ are $\mathbb{G}^{\textrm{(0)}}_{xx} = \mathbb{G}^{\textrm{(0)}}_{yy}$ and $\mathbb{G}^{\textrm{(0)}}_{zz}$ one can show that $\Gamma_{21} = \Gamma^{(0)}$, as it should be.

In an inhomogeneous environment Eq. (\ref{FreeSpaceGreenFunction}) does not give 
the full Green function of the problem. In this case the dyadic Green function 
has to be modified in order to satisfy the 
electromagnetic field boundary conditions and to take into account scattering 
owing to neighboring objects. Particularly,  for an emitter close to a 
half-space presenting a flat interface at \linebreak $z = 0$, one can write $ 
\mathbb{G} ({\bf r}, {\bf r^{\prime}}; \omega ) = \mathbb{G}^{(0)} ({\bf r}, 
{\bf r^{\prime}};\omega ) + \mathbb{G}^{(S)} ({\bf r}, {\bf r^{\prime}};\omega 
)$, where \cite{novotnybook}
%
\begin{equation}
	{\mathbb G}^{(S)} ({\bf r}, {\bf r^{\prime}}; \omega) = \frac{i}{2}  \int \frac{\mathbb{R} \, e^{i \mathbf{k_{\parallel}} \cdot (\mathbf{ r} - \mathbf{ r^{\prime}})}  e^{i k_{z0} (z+z^\prime)}}{k_{z0}} \frac{d^2 \mathbf{k_{\parallel}}}{(2\pi)^2}\, ,
\end{equation}
is the Green function associated to the electromagnetic field generated by the oscillating dipole source and scattered (reflected) by the semi-infinite medium. $\mathbb{R}$ is the half-space reflection matrix given by
\begin{equation}
\label{ReflectionMatrix}
\mathbb{R}\ \ =\!\!\! \sum_{i\, , \ j = \{\textrm{TE, TM}\}}\!\!\!\!\!\!\!\!
r^{\textrm{i, j}}   \mbox{{\mathversion{bold}${\epsilon}$}}_{\textrm{i}}^{+} 
\otimes \mbox{{\mathversion{bold}${\epsilon}$}}_{\textrm{j}}^{-}\, ,
\end{equation}
where $r^{\textrm{i, j}}$ ($i, j =$ TE, TM) corresponds to the 
reflection coefficient for incoming  j-polarized light 
that is reflected as an i-polarized wave.

Given this system geometry, one can decompose the SE rate in two contributions 
$\Gamma_{21} = \Gamma_{\perp} + \Gamma_{\parallel} $. Here, $ \Gamma_{\perp}$ 
($ \Gamma_{\parallel}$) expresses the decay rate contribution due to the 
transition dipole moment component perpendicular (parallel) to the vacuum-medium 
interface. Plugging Eqs. 
(\ref{FreeSpaceGreenFunction})-(\ref{ReflectionMatrix}) into Eq. 
(\ref{TaxaEmissaoEspontanea2}) one can show that \cite{kortkamp2015}
\begin{align}
	\dfrac{\Gamma_{\perp}}{\Gamma^{(0)}}  =& \dfrac{d_{z}^2}{|{\bf d}_{21}|^2} \bigg\{ 1 + \dfrac{3}{2} \int_0^{k_0}\dfrac{k_{\parallel}^3}{k_{0}^3\,\xi} \textrm{Re}\left[ r^{\textrm{TM, TM}} \, e^{2i\xi z} \right] \, d k_{\parallel} \nonumber \\ 
													&+  \dfrac{3}{2}\int_{k_0}^{\infty} \dfrac{k_{\parallel}^3}{k_{0}^3\,\zeta} \, e^{- 2 \zeta z } \textrm{Im}\left[ r^{\textrm{TM, TM}} \right] \, d k_{\parallel} \bigg\}\, , 
\label{GammaPerp}
\end{align}
and
\begin{align}
	\dfrac{\Gamma_{\parallel}}{\Gamma^{(0)}} =& \dfrac{d_{\parallel}^2}{|{\bf d}_{21}|^2} \bigg\{ 1 + \dfrac{3}{4} \int_0^{k_0}\dfrac{k_{\parallel}}{k_{0}^3\,\xi} \textrm{Re}\left[\left(k_0^2 r^{\textrm{TE, TE}}  \right. \right. \nonumber \\ 
						 &  \left. \left. - \, \xi^2 \, r^{\textrm{TM, TM}}\right) e^{2i\xi z}\right]  dk_{\parallel} + \dfrac{3}{4} \int_{k_0}^{\infty}\dfrac{k_{\parallel}}{k_{0}^3\,\zeta}\, \textrm{Im}\left[k_0^2 \, r^{\textrm{TE, TE}} \right. \nonumber \\
						 & \left. + \, \zeta^2 \, r^{\textrm{TM, TM}}\right] e^{-2\zeta z} dk_{\parallel} \bigg\}\, , 
\label{GammaPar}
\end{align}
where  $\xi = \sqrt{k_0^2 - k_{\parallel}^2}$ and $\zeta = \sqrt{k_{\parallel}^2 
- k_0^2}$.  In the cases we  consider $r^{\textrm{TE, TE}}$ and 
$r^{\textrm{TM, TM}}$ will be given by the usual Fresnel reflection coefficients 
for a flat interface between vacuum and an homogeneous medium, namely 
\cite{novotnybook}
\begin{eqnarray}
\label{FresnelCoefficients}
r^{\textrm{TE, TE}} = \dfrac{{k_z}_0 - {k_z}_1}{{k_z}_0 + {k_z}_1}\, ,\ \ \ 
r^{\textrm{TM, TM}}=\dfrac{\varepsilon_e{k_z}_0- {k_z}_1}{\varepsilon_e{k_z}_0 +{k_z}_1}\, ,
\end{eqnarray}
where $\varepsilon_e$ is the effective dielectric constant of the substrate (see 
next section) and $k_{z1} = \sqrt{\varepsilon_ek_0^2-k_{\parallel}^2}$. Finally, note that for an isotropic atom we have $d_{z}^2/|{\bf 
d}_{12}|^2 = 1/3$ and $d_{\parallel}^2/|{\bf d}_{12}|^2 = 2/3$.

\subsection{Effective Medium Theory}
\label{EMT}

Effective medium theories allow one to 
construct an effective dielectric constant $\varepsilon_e$ 
of a composite medium as a function of its 
constituents' properties (dielectric constants and shapes) as well as of the 
fractional volumes characterizing the mixture~\cite{Lagarkov1996, 
Brouers1986,Goncharenko2004,Choy1999}.

One of the most important and successful 
effective medium approaches is the Bruggeman Effective Medium Theory (BEMT), 
which is the simplest analytical model that predicts an insulator-metal 
transition at a critical concentration of metallic particles in the dielectric 
host \cite{Choy1999,Sahimi1993}. BEMT treats the dielectric host medium and the 
metallic inclusions symmetrically, and it is based on the following assumptions: 
$(i)$ the grains are randomly oriented spheroidal 
particles, and $(ii)$ they are embedded in an homogeneous effective medium of 
dielectric constant $\varepsilon_e$  that will be determined self-consistently 
\cite{Sahimi1993}. In this work we consider spheroidal inclusions whose 
geometry is characterized by the depolarization factor $0\leq L\leq1$. 
Explicit expressions of $L$ in terms of the eccentricity $e$ of the spheroid 
are \cite{kortkamp2014}
\begin{eqnarray}
\!\!\!\!\!\!\!\!\!\!\!\!\label{DepoalrizationFactor}
L = \left\{ \!
\begin{tabular}{c}
$\dfrac{1-e^2}{2e^3}\left[\ln\left(\dfrac{1+e}{1-e}\right) -2e \right]\!,$ prolate spheroid\\ \\
$\dfrac{1+e^2}{e^3}\left[e-\arctan(e) \right]\!,$ oblate spheroid\\
\end{tabular}
\right.
\end{eqnarray}

Within the BEMT $\varepsilon_{e}$ is computed by 
demanding that the average over all directions of the 
scattered Poynting vector vanish when the system is illuminated 
by a monochromatic wave with wavelength (both in vacuum and inside the medium) 
much larger than the size of the inclusions. In this case, the effective 
permittivity satisfies the following equation~\cite{Lagarkov1996, 
Brouers1986,Goncharenko2004,Choy1999, kortkamp2014},
\begin{eqnarray}
(\!\!\!&1&-f)\!\left\{\dfrac{\varepsilon_{hm} - \varepsilon_{e}}{\varepsilon_{e} + L(\varepsilon_{hm}-\varepsilon_{e})}
+ \dfrac{4(\varepsilon_{hm} - \varepsilon_{e})}{2\varepsilon_{e} + (1-L)(\varepsilon_{hm}-\varepsilon_{e})}\right\}\cr
\!\!\!\!\!\!&+&f\!\left\{\dfrac{\varepsilon_{i} - \varepsilon_{e}}{\varepsilon_{e}\!\! +\!\! L(\varepsilon_{i}-\varepsilon_{e})}
+ \dfrac{4(\varepsilon_{i} - \varepsilon_{e})}{2\varepsilon_{e} + (1-L)(\varepsilon_{i}-\varepsilon_{e})}\right\}=0 \, \label{BEMT} ,
\end{eqnarray}
where $\varepsilon_i$, $\varepsilon_{hm}$ are the dielectric constants of the metallic inclusions and host matrix, respectively, and $f$ ($0\leq f \leq 1$) is the volume filling factor for the metallic inclusions. Equation (\ref{BEMT}) has several roots but only the one with $\textrm{Im}(\varepsilon_e) \geq 0$ is physical since we are assuming passive materials ({\it i. e.}, no optical gain).

The percolation threshold $f_c$ corresponds to a critical value of the filling factor for which 
the composite medium undergoes an insulator-conductor transition, 
thereby exhibiting a dramatic change in 
its electrical and optical 
properties~\cite{Goncharenko2004,Brouers1986,Lagarkov1996,Sahimi1993, 
kortkamp2014,Sarychev2000}. This critical filling factor is calculated by taking 
the quasi-static limit ($\omega \rightarrow 0$) in Eq. (\ref{BEMT}). In this 
limit, $\varepsilon_i \gg \varepsilon_{hm}$, $\textrm{Im}[\varepsilon_i] \gg 
\textrm{Re}[\varepsilon_i]$, and $\textrm{Im}[\varepsilon_{hm}] \ll 
\textrm{Re}[\varepsilon_{hm}]$ provided the host medium does not have a 
resonance near $\omega = 0$. Consequently, $\varepsilon_i$ ($\varepsilon_{hm}$) 
may be approximated by a pure imaginary (real) function. Besides, if $f<f_c$ 
$(f\geq f_c)$ the effective medium behaves as a dielectric-like (metal-like) 
material so that $\textrm{Re}[\varepsilon_e] > 0$ ($\textrm{Re}[\varepsilon_e] 
<0$)  in the low frequency regime. Hence, the critical threshold filling factor 
can be obtained by the condition $\textrm{Re}[\varepsilon_e] = 0$.  For 
spheroidal inclusions the BEMT predicts that the percolation transition occurs 
at~\cite{Brouers1986,Lagarkov1996,Goncharenko2004,kortkamp2014}
\begin{equation}
f_{c}^B(L) = \frac{L(5-3L)}{(1+ 9L)}.
\label{bruggeman}
\end{equation}

In order to test the robustness of our results with respect to specific features of a given effective 
medium theory,  we shall consider an alternative 
homogenization technique  proposed by Lagarkov and Sarychev in 
Ref.~\cite{Lagarkov1996} as well. The 
Lagarkov-Sarychev approach is known to give more accurate results for $f_{c}$ 
than the BEMT in the regime of small $L$ ($L \ll 1 $);  the  critical filling 
factor within the Lagarkov-Sarychev effective medium theory is\cite{Lagarkov1996}
\begin{eqnarray}
f_c^{LS}(L) = \dfrac{9L(1-L)}{2+15L-9L^2}\, .
\label{lagarkov}
\end{eqnarray}

In the following section the SE rate of an emitter close to 
a semi-infinite composite medium will be computed by means of the 
effective medium approaches described above. The dielectric functions 
of the metallic inclusions $\varepsilon_i$ and of the dielectric host-medium 
$\varepsilon_{hm}$ are
\begin{eqnarray}
\varepsilon_{i} (\omega) &=& 1 - \frac{\omega_{pi}^2}{\omega^2 + i \, 
\gamma_i \omega}\, , \\
\varepsilon_{hm} (\omega) &=& 1 + \sum_{j} \frac{{\omega_{pj}^{hm}}^2}{\omega_{Rj}^2 \,-\,\omega^2 + i \, \omega \, \Gamma_{j}}\, ,
\end{eqnarray}
where $\omega_{pi}\ ({\omega_{pj}^{hm}})$ and $\gamma_i\ (\Gamma_j)$ are, 
respectively, the plasma frequency (oscillating strengths) and the inverse of 
the relaxation time(s) of the 
metallic inclusions (host medium). The value of 
these parameters for the metals (Au, Cu, Ti, Ag) and 
dielectrics (polystyrene) considered were extracted from Refs.~\cite{Ordal1985, Hough80, Comment1}.

\section{Results and Discussions}
\begin{figure}
\centering
\includegraphics[scale=0.38]{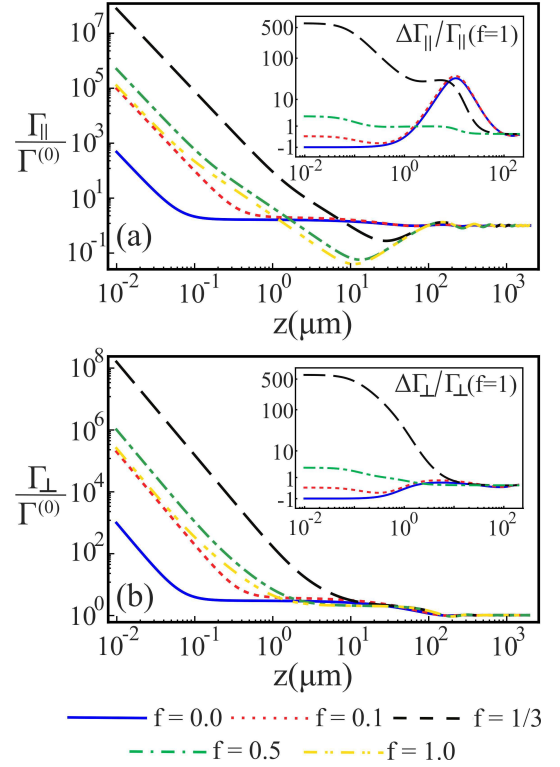}
\caption{Spontaneous emission rate dependence with the distance $z$ to the semi-infinite composite medium (BEMT) for parallel (a) and perpendicular (b) transition dipole orientations. In both panels the metallic inclusions are made of gold and $L = 1/3$ (spheres). The insets display the  SE rates relative to the $f=1$ case, i. e., when the emitter is in presence of an homogeneous gold substrate, for the same values of $f$ as before. }
\label{Figure2}
\end{figure}

In Fig.~\ref{Figure2} the SE rate is calculated as a function of the distance 
between the emitter and the semi-infinite medium made of spherical ($L = 1/3$) 
gold inclusions embedded in a polystyrene host matrix for different values of 
the filling fractions $f$. Within BEMT the dielectric constant of the composite 
medium presents a dielectric-like (metal-like) response for $f<1/3$ ($f>1/3$) 
\cite{Sahimi1993}. The 
emitter is assumed to be a Caesium (Cs) atom with transition wavelength 
$\lambda = 450 \ \mu$m, that is, in the THz frequency range. It 
is clear that the composite media may greatly 
enhance the SE rate when compared to the homogeneous cases $f = 0$ and $f = 1$ 
for both transition electric dipole parallel and perpendicular to the flat 
interface. 
Indeed, in the presence of the composite media the emitter's decay rate 
may be five to six orders of magnitude times larger than its value in 
free space for distances $z \sim 100$ nm, as it can be seen in Figs.~\ref{Figure2}a  and \ref{Figure2}b for 
$\Gamma_{\parallel}$  and $\Gamma_{\perp}$, respectively. Figure~\ref{Figure2} 
also reveals that the transition from far- to near-field effects on the 
emitter's lifetime can be tuned by the filling factor 
$f$. Interestingly, for $f = f_c^B = 1/3$ 
near-field effects become relevant even for distances of the order of $z \sim 1 \mu$m. Similar qualitative 
results hold for the perpendicular configuration, even though $f$ seems to play 
a less prominent role in the near-to-far-field transition distance range. 

It should be noticed that in the far-field regime the dependence of the SE rate 
on $f$ is very weak. For large distances  $\Gamma_{\parallel}$  and 
$\Gamma_{\perp}$ can be approximated by the first integrals in Eqs. 
(\ref{GammaPerp}) and (\ref{GammaPar}) with the main contribution originating 
from electromagnetic modes with $k_{z0} = \xi \simeq 0$ (due to the oscillatory 
behavior of $e^{2i\xi z}$). An expansion of the reflection 
coefficients around $k_{z0} = 0$ shows that $r^{\textrm{TE, TE}} \simeq 
r^{\textrm{TM, TM}} \simeq -1 + {\cal{O}}(\xi/k_0)$ and, hence, the dominant 
contribution to the emitter's decay rate in the far-field does not 
carry information about the electromagnetic properties of the substrate. On the 
other hand, the metal concentration  strongly affects the Purcell effect for 
distances $z\lesssim 1\ \mu$m. For such distances light emission is more 
affected by electromagnetic evanescent modes ($k_{\parallel}>k_0$) that exist 
only close to the air-substrate interface. Particularly, in the extreme 
near-field regime an approximate analytic expression for the SE rate can be 
obtained by taking the quasi-static limit ($c\to\infty$)\linebreak  in Eqs. 
(\ref{GammaPerp}) and (\ref{GammaPar}), 
\begin{equation}
\dfrac{\Gamma_{\perp}}{\Gamma^{(0)}} \simeq 2 \dfrac{\Gamma_{\parallel}}{\Gamma^{(0)}} \simeq \dfrac{3}{4} \dfrac{1}{z^3} \dfrac{\textrm{Im}[\varepsilon_e]}{|\varepsilon_e+1|^2}\, .
\label{SENF}
\end{equation}
Note that the $z^{-3}$ distance scalling-law for bulk materials is 
rederived regardless of the optical characteristics 
of the substrate. On the one hand, for low values of 
$f$ the decay rate dynamics is governed by the (small) losses in the dielectric 
host since $\Gamma_{\perp, \parallel}$ are proportional to 
Im$[\varepsilon_{e}]\simeq$ Im$[\varepsilon_{hm}]\ll1$. The SE rate increases as 
small amounts of metallic inclusions are added to the host matrix due to 
enhancement of absorption processes in the substrate. On the other 
hand, large concentrations of metal lead to SE rates 
proportional to 1/Im$[\varepsilon_{e}]\simeq$ 1/Im$[\varepsilon_{i}]\ll1$. Based 
on this analysis, its clear that the emitter's lifetime will be greatly modified 
by $f$, as seen in \linebreak Figs. \ref{Figure2}a and b for $z\lesssim 1\ \mu$m.

\begin{figure}
\centering
\includegraphics[scale=0.35]{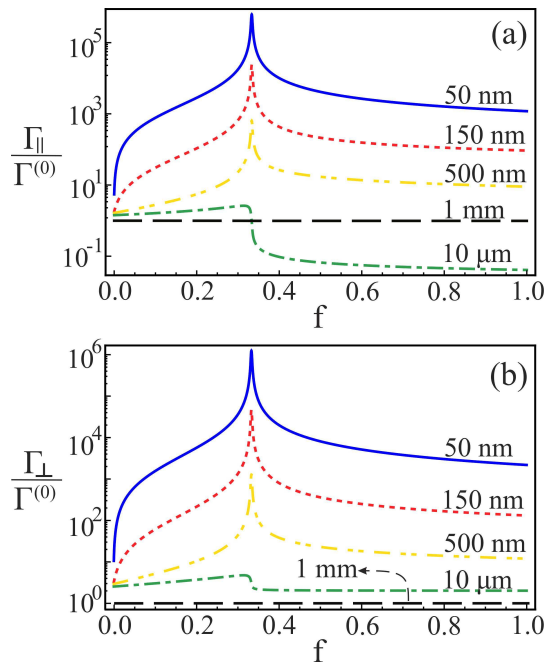}
\caption{Spontaneous emission rate as a function of the volume filling factor $f$ for parallel $\Gamma_{\parallel}/ \Gamma^{(0)}$ (a) and perpendicular $\Gamma_{\perp}/ \Gamma^{(0)}$ (b) cases. Results were computed for several $z$-distances between the emitter and the semi-infinite medium. All other parameters are the same as in Fig. \ref{Figure2}. Both panels reveal that maximum enhancement in the SE rate occurs at the insulator metal transition $f_c^B = 1/3$.}
\label{Figure3}
\end{figure}

The fact that in the near-field regime $\Gamma_{\perp, \parallel}$ initially 
grows with Im$[\varepsilon_{e}]$ ($f\ll1$) and then decays with 
1/Im$[\varepsilon_{e}]$ ($f\simeq1$)  suggests that the decay rate should 
present a peak at some critical filling factor. Remarkably, the SE rate reaches 
its maximum value at $f_{c}^B = 1/3$ (see Fig. \ref{Figure3}), which precisely 
corresponds to the percolation transition threshold predicted by the BEMT for 
spherical inclusions.  At $f_c^B$ the value of the SE can be more than two 
orders of magnitude larger than its values for other inclusions' concentrations 
($f \neq f_c^B$) and for distances $z  \lesssim 1$ $\mu$m. In the 
insets of Fig.~\ref{Figure2}, the relative variation of the SE rate with respect 
to its value in the presence of an homogeneous gold semi-infinite medium 
($f=1$),

\begin{equation}
\frac{\Delta \Gamma_{\perp,\parallel}}{\Gamma_{\perp,\parallel}(f=1)} 
\equiv  \frac{\Gamma_{\perp,\parallel} - 
\Gamma_{\perp,\parallel}(f=1)}{\Gamma_{\perp,\parallel}(f=1)},
\end{equation}
is calculated as a function of $z$. In both parallel and 
perpendicular cases, $\Delta 
\Gamma_{\perp,\parallel}/\Gamma_{\perp,\parallel}(f=1)$ is largely enhanced at 
the percolation transition $f_c^B$, specially for shorter distances $z  \lesssim 
1$ $\mu$m. This relative variation can be as impressive as 500 at $f_c^B$, 
unambiguously demonstrating that composite media can largely outperform 
homogeneous media, when it comes to modify and tune the SE rate.     

In order to further investigate the dependence of the SE rate on the filling 
fraction $f$, in Fig.~\ref{Figure3} the behavior of 
$\Gamma_{\perp}$ and $\Gamma_{\parallel}$ as a fuction of $f$ is depicted for 
different distances $z$ between the emitter and the composite medium. Figure 
\ref{Figure3} shows that for distances $z  \lesssim 1\ 
\mu$m the decay rate reaches its maximal value exactly at the 
percolation threshold $f_{c}^B$, in both parallel and perpendicular 
configurations. For distances smaller than $1\ \mu$m the results of Fig. 
\ref{Figure3} are well described by Eq. (\ref{SENF}). As a consequence, 
in this 
regime the enhancement in the Purcell effect at $f_c^B$ relative to 
the homogeneous gold semi-infinite medium is distance-independent, 
$$
\frac{\Gamma_{\perp,\parallel}(f = 
f_c^B)}{\Gamma_{\perp,\parallel}(f = 1)} \simeq 
\frac{{\rm Im}[\varepsilon_e]}{{\rm Im}[\varepsilon_i]} 
\frac{|\varepsilon_i+1|^2}{|\varepsilon_e+1|^2} .
$$  
For other distances $z  \lesssim 10\ \mu$m the enhancement at $f_{c}^B$ is of one order of 
magnitude or less. Figure \ref{Figure3} also emphasizes the importance of 
near-field effects, as the SE rate enhancement becomes small for distances $z 
\gtrsim 10$ $\mu$m and completely disappears at $z \sim \unit[1]{m m}$.

It is interesting to comment that a large enhancement of heat transfer between 
composite bodies at the percolation threshold also occurs due to near-field 
effects~\cite{kortkamp2014}. Here the role of the near-field is similar and it 
helps one to qualitatively understand the physical origin of the SE rate 
enhancement at $f_{c}$. Indeed, the physical explanation of these two distinct 
phenomena (SE decay rate and near heat field transfer) in the presence of 
composite media are intrinsically related to the universal properties of the 
percolation phase transition, in particular the enhanced and scale invariant 
current and electric field fluctuations that take place close to the percolation 
critical point~\cite{Sarychev2000}. These enhanced electric field fluctuations 
modify the structure of the electromagnetic modes, and hence show up in the 
LDOS. For composite media around the percolation threshold, extremely localized 
and subwavelength confined resonant plasmon excitations occur, leading to the 
formation of giant spatial fluctuations of the electromagnetic field intensity 
(``hot spots'')~\cite{Sarychev2000}.  As the existence of localized modes has 
been demonstrated to strongly amplify the LDOS~\cite{krachmalnicoff2010}, we 
conclude that the SE rate should be enhanced at the insulator-metal transition 
as well. These arguments qualitatively explain the results reported in 
Figs.~\ref{Figure2} and~\ref{Figure3}.   
\begin{figure}
\centering
\includegraphics[scale=0.21]{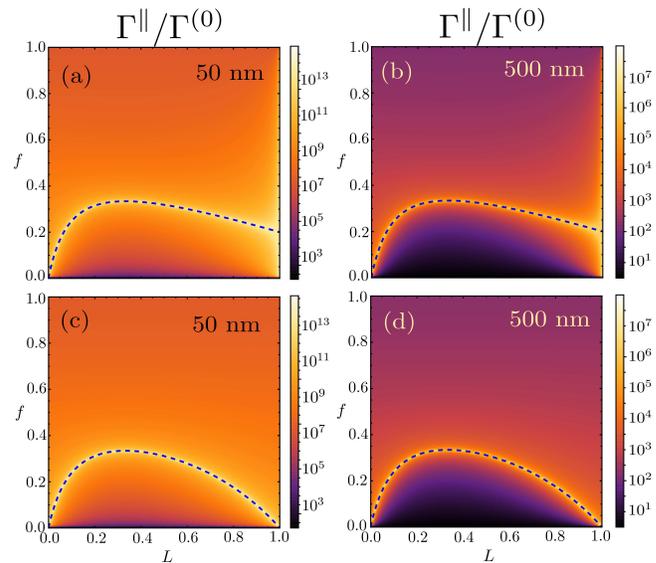}
\caption{Density plots of $\Gamma_{\parallel}/ \Gamma^{(0)}$ as a function of both filling factor $f$ and the depolarization factor $L$ for $z = 50$ nm (a) and (c), and $z = 500$ nm (b) and (d). Panels (a) and (b) correspond to calculations using the Bruggeman effective medium theory whereas (c) and (d) to the Lagarkov-Sarychev approach. The dashed line in each plot shows the percolation transition curve as predicted by Eqs. ~\ref{bruggeman} and \ref{lagarkov}. All other parameters are the same as in Fig. \ref{Figure2}.} 
\label{Figure4}
\end{figure}

 The influence of the various possible inclusions shapes in the SE rate in shown 
in Fig.~\ref{Figure4}, where $\Gamma_{\parallel}$  is calculated as a function 
of both the filling fraction $f$ and the depolarization factor $L$ for $z = 50$ 
nm [(a) and (c)] and $z = 500$ nm [(b) and (d)]. In panels (a) and (b) the 
effective 
dielectric constant of the substrate was obtained using BEMT whereas (c) and (d) 
correspond to calculations involving the Lagarkov-Sarychev approach. We note 
from these graphics that for each value of $L$ there exist an optimum filling 
factor $f$ that maximises the decay rate (brightest regions in the plots). 
Interestingly, the position of  the peak of the SE rate in Fig. \ref{Figure4} is 
perfectly described by the percolation curves given in Eqs. (\ref{bruggeman}) 
and (\ref{lagarkov}), as shown by the dashed lines in the plots. These results 
demonstrate the robustness of our findings against variations of the shape of 
the  inclusions as well as changes in effective theory used to model the 
electric permittivity $\varepsilon_{e}$ of the composite medium. We 
checked that these conclusions are  valid for both parallel 
($\Gamma_{\parallel}$) and perpendicular ($\Gamma_{\perp}$) configurations and 
apply for all distances $z\lesssim 10\ \mu$m.  We have also verified that our 
results are not qualitatively modified by changing the material that constitute 
the metal inclusions. Indeed, in Table ~\ref{Metais}, we show the ratio between 
$\Gamma_{\parallel}$ 
at the percolation threshold ($f=f_c$) and its value for a full metallic 
semi-infinite medium ($f = 1$) when the substrate is composed with different 
metal inclusions. In the table 
$L=0.1$ corresponds to needle-like inclusions (eccentricity $e \simeq 0.95$) and 
 $L=1/3$ to spherical ones ($e = 0$). The enhancement in the SE 
rate$\Gamma_{\parallel}$ is at least two orders of magnitude, regardless of the 
metal and the effective medium theory considered. We emphasize that the same 
effect occurs for the $\Gamma_{\perp}$ rate.  These results provide evidence 
that our findings should hold even beyond the effective medium approximation. 
The differences in the results obtained by means of the BEMT and the 
Lagarkov-Sarychev model for $L=0.1$ are due to the distinct assumptions made 
about the host dielectric medium (see Sec. \ref{EMT} and Refs. 
\cite{Brouers1986,Goncharenko2004,Sahimi1993,Sarychev2000}). 

\begin{table}
	\centering
	\begin{tabular}{c|c|c|c|c}
	& \multicolumn{2}{c}{Bruggeman} &  \multicolumn{2}{|c}{Lagarkov-Sarychev} \\ \cline{2-5}
	&$L=0.1$&$L=1/3$&$L=0.1$& $L=1/3$ \\ \hline
    Au & 794  & 524 & 116 & 524 \\
    Cu & 1099 & 726 & 104 & 726 \\
    Ti & 190  & 127 & 127 & 127 \\
    Ag & 923  & 610 & 108 & 610
\end{tabular}
\caption{Values for the ratio 
$\Gamma_{\parallel}(f=f_c)/\Gamma_{\parallel}(f=1)$ for different metallic 
inclusions. All values were computed by considering a Polystyrene host medium, 
an emitter-substrate distance of $\unit[50]{nm}$, and emission wavelength of 
$450 \mu m$.}
\label{Metais}
\end{table}

\begin{figure}
\centering
\includegraphics[scale=0.35]{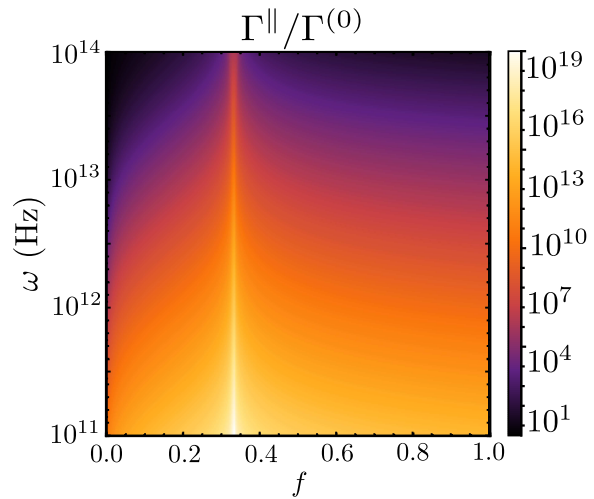}
\caption{The two-dimensional plot displays the ratio $\Gamma_{\parallel}/ \Gamma^{(0)}$ using the BEMT as a function of the filling factor $f$ and the quantum emitter frequency $\omega$.  The enhancement of the SE rate reaches its maximum precisely at percolation transition $f=f_c^B = 1/3$ (for spherical inclusions). A similar behavior has been verified for the perpendicular case $\Gamma_{\perp}/ \Gamma^{(0)}$.}
\label{Figure5}
\end{figure}

We should mention that our results apply for a broad range of transition 
frequencies and could be tested using quantum emitters working from THz to 
near-infrared. In Fig. \ref{Figure5}, the SE rate in the parallel configuration 
is calculated using BEMT as a function of both transition frequency $\omega$ and 
volume filling factor $f$
considering spherical gold inclusions ($L = 1/3$) at $z = \unit[50]{nm}$ 
distance between the emitter and the composite media. The maximum emission 
always take place at the percolation threshold $f_c^B$. At larger frequencies 
the insulator-metal transition effect on the emitter's lifetime becomes weaker 
and a broadening of the emission peak occurs. 
This behavior is due to the fact that, for larger and larger 
frequencies, the 
distinction between dielectrics and conductors becomes less and less 
pronounced, thus making the percolation transition less dramatic. We have 
verified that similar results hold in the perpendicular configuration; they also 
apply within the Lagarkov-Sarychev model.

\begin{figure}
\centering
\includegraphics[scale=0.42]{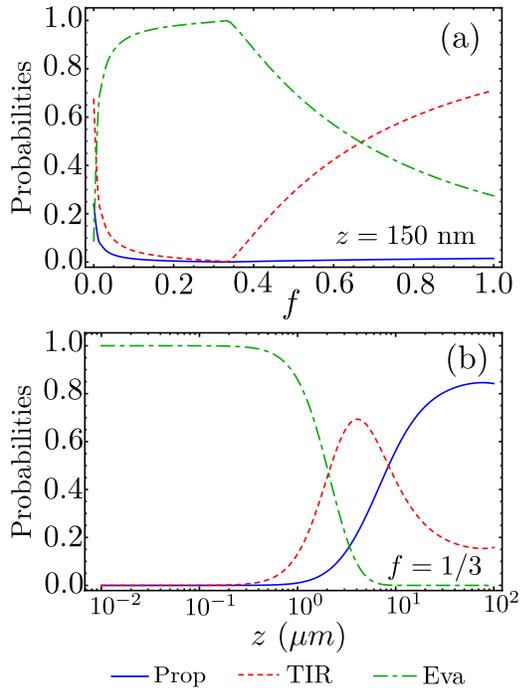}
	\caption{The decay channel probability of the quantum emission in the presence of the effective media (BEMT) with spherical inclusions ($L =1/3$) is displayed as a function of both the filling factor $f$ at a fixed distance $z = \unit[150]{nm}$ between the atom and the media (a) and as a function of $z$ at the percolation threshold $f = 1/3$ (b).}
\label{Figure6}
\end{figure}

In Fig.~\ref{Figure6}, we investigate the role of the different decay 
channels in the SE rate. The decay probability of the quantum emission is shown 
in terms of propagating (Prop), totally internal reflected (TIR) or evanescent 
modes (Eva) as a function of the filling factor $f$ and the distance $z$ between 
the atom and the semi-infinite medium. The probabilities are computed as the 
ratio between the partial and the total SE rate. The partial contribution of 
these modes to decay rates can be expressed as~\cite{kortkamp2015}
\begin{align}
	\dfrac{\Gamma_{\perp}^{\rm{Prop}}}{\Gamma^{(0)}} (z) =& 1 + \dfrac{3}{2} \int_0^{k_0}\dfrac{k_{\parallel}^3}{k_{0}^3\,\xi} \textrm{Re}\left[ r^{\textrm{TM, TM}} \, e^{2i\xi z} \right] \, d k_{\parallel} \nonumber \\
	\dfrac{\Gamma_{\perp}^{\rm{TIR}}}{\Gamma^{(0)}} (z) =& \dfrac{3}{2} 
\int_{k_0}^{n_e k_0} 
\dfrac{k_{\parallel}^3}{k_{0}^3\,\zeta} \, e^{- 2 \zeta z } \textrm{Im}\left[ 
r^{\textrm{TM, TM}} \right] \, d k_{\parallel}  \nonumber \\
	\dfrac{\Gamma_{\perp}^{\rm{Eva}}}{\Gamma^{(0)}} (z) =& \dfrac{3}{2} 
\int_{n_e k_0}^\infty 
\dfrac{k_{\parallel}^3}{k_{0}^3\,\zeta} \, e^{- 2 \zeta z } \textrm{Im}\left[ 
r^{\textrm{TM, TM}} \right] \, d k_{\parallel} ,
\end{align}
where we defined the medium index of refraction $n_e = \textrm{Re} \sqrt{\varepsilon_e/\varepsilon_0}$. In Fig. 6(a) we see clearly that for $f \leq f_c$, the contribution of the evanescent 
modes quickly rises and dominates the decay process. Despite the fact that the medium behaves effectively as a dielectric in this regime, it is the dissipation in the metallic inclusions that actually gives rise to such a dominance. Once we step into the $f>f_c$ region, we see that evanescent and TIR modes progressively swap roles, and the latter becomes the most important decay channel.  This happens because $n_e$ grows steadily as a function of $f$ in the metallic regime, so there are more TIR modes available as the filling factor is increased. In addition, in Fig. 6(b) we show the different decay pathways right at the percolation threshold \cite{Comment2}, as a function of distance. For short distances, the enhancement in SE is due mainly to 
the evanescent contribution, meaning that the energy associated with the decay 
is (with very high probability) absorbed by the half-space \cite{FordWeber}.
As the distance increases, the contribution of evanescent modes 
decreases and propagating and TIR modes become more important. 

\begin{figure}
\centering
\includegraphics[scale=0.35]{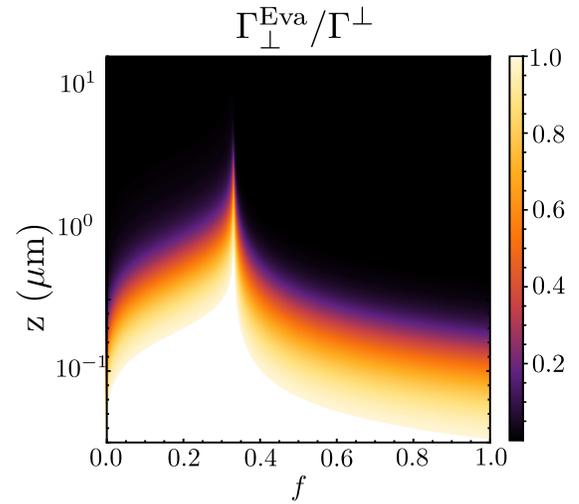}
\caption{The quantum emission into evanescent modes is displayed as a function of 
both the distance between the emitter and the media $z$ and the filling factor 
$f$.}
\label{Figure7}
\end{figure}

 In Fig. \ref{Figure7}, the 
contribution to the SE rate due to evanescent modes is shown as a function of 
both $f$ and $z$. It can be noted that, for $z \lesssim \unit[100]{nm} $, the 
probability of decaying in a evanescent mode is larger than $60\%$ (and even 
larger when the media behaves like dielectric $f < f_c^B$). We should also 
stress out that at the percolation $f = f_c^B =1/3$ (for spherical inclusions) 
the contribution of evanescent modes are relevant even for larger distances of 
about $z \lesssim  \unit[1]{\mu m}$.

\section{Conclusions}

In conclusion, we have investigated the spontaneous emission rate of a two-level 
atom in the vicinities of a semi-infinite medium composed of randomly dispersed, 
arbitrary shaped gold inclusions embedded in a polystyrene host matrix.  Using 
effective medium theories to describe the 
electromagnetic properties of the composite medium, we 
demonstrate that the presence of composite media is responsible for a great 
enhancement of the SE rate relative to the homogeneous semi-infinite medium 
case. We find that this enhancement in the SE rate is maximal at the percolation 
critical point for the composite medium, where it can be as impressive as two 
orders of magnitude. The enhancement in the spontaneous emission rate is more 
pronounced at small distances between the emitter and the composite medium, 
unveiling the crucial role of near field effects. In addition, we show that our 
results are robust against material losses, to changes in the shape of 
inclusions and materials, for a broad range of transition frequency, and apply 
for different effective medium theories. We also investigate the contribution of 
different decaying channels in the SE rate. We hope that our findings could 
guide the design of composite media aiming at tailoring and optimizing the decay 
rate of quantum emitters.   

\section*{Acknowledgements}
We thank E. C. Marino and M. Hippert for useful discussions. D.S., F.S.S.R. and C. F. acknowledge CAPES, CNPq, and FAPERJ for financing this research. W.K.-K. thanks LANL LDRD project for financial support. F.A.P. thanks the hospitality of the Optoelectronics Research Centre and Centre for Photonic Metamaterials, University of Southampton, where part of this work has been done, and CAPES for funding his visit. F.A.P. acknowledges CAPES (Grant No. BEX 1497/14-6) and CNPq (Grant No. 303286/2013-0)


\begin{thebibliography}{99}
\bibitem{Comment1} In Ref. [32], $\Gamma_1 = \Gamma_2 = 0$. Here, however, we decided to include small ohmic losses so as to
make the model more realistic.
\bibitem{Comment2} The percolation transition is defined in the limit $\omega \rightarrow 0$, and we, strictly speaking, are working at a finite frequency ($\omega_0 \approx 0.8$ THz). That makes critical filling factor $f_c = 1/3$ to be slightly off (we have actually $f_c \approx 0.336$), but this has absolutely no effect in our results.
\end{thebibliography}
\end{document}